# WIKI-INDEX OF AUTHORS' POPULARITY

D.V. Lande, V.B. Andrushchenko, I.V. Balagura

Institute for Information Recording of NAS of Ukraine, Kiev

**Abstract**

*The new index of the author's popularity estimation is represented in the paper. The index is calculated on the basis of Wikipedia encyclopedia analysis (Wikipedia Index – WI ). Unlike the conventional existed citation indices, the suggested mark allows to evaluate not only the popularity of the author, as it can be done by means of calculating the general citation number or by the Hirsch index, which is often used to measure the author's research rate. The index gives an opportunity to estimate the author's popularity, his/her influence within the sought-after area "knowledge area" in the Internet – in the Wikipedia.*

*The suggested index is supposed to be calculated in frames of the subject domain, and it, on the one hand, avoids the mistaken computation of the homonyms, and on the other hand – provides the entirety of the subject area.*

*There are proposed algorithms and the technique of the Wikipedia Index calculation through the network encyclopedia sounding, the exemplified calculations of the index for the prominent researchers, and also the methods of the information networks formation – models of the subject domains by the automatic monitoring and networks information reference resources analysis.*

*The considered in the paper notion network corresponds the terms-heads of the Wikipedia articles.*

**Introduction**

Today scientometric mostly uses several indices, according to which the scientists' rate and their impact on science and society are calculated. Thus, the simplest index is the number author's publications. It is clear that this index does not depict the qualitative parameters that are better reflected in another index - the number of citations. This index doesn't illustrate the overall performance of the author because the author of just the one, but very important work may exceed this indicator in comparison with scientists who regularly publish their results.

In 2005 the physician Jorge E. Hirsch from the California University established the most popular index – Hirsch Index.

The principle of its calculation is quite simple, while it combines the advantages of the first and second approaches. The index calculation is based on the distribution of citations of the work of researcher. According to Hirsch scientist has index h, if h of his Np papers cited at least h times each, while both articles remaining (Np - h) quoted no more than h times each. This index gained the support and is used in such scientometric systems as Scopus, Web of Science, and Google Scholar Citations.

At the same time this indicator, which is focused on the scientific importance, significance of the author, not quite fully reflects the overall importance of the results that he/she received. For such an assessment it is appropriate to use non-fiction and open access systems. As one of the approaches to solve this problem, the authors proposed methodology for calculating the new index - the Wiki-index of authors' popularity.

This index can appear an important tool in combination and with other indices can provide a complete picture of influential scientific achievements of the author, not only in the research community, but the overall impact on the formation of perspective and fully understanding of research information by the users.

The network service Wikipedia - the largest and most democratic Internet encyclopedia is considered, access to which does not presume subscription and furthermore the system is available for download in full.

Today Wikipedia (https://www.wikipedia.org/) is the most visited site in the Internet, and one of the most popular encyclopedic resources covering all the disciplines, it provides answers to the most search engines queries. At this time only the English version of Wikipedia contains more than 5 million articles (German - more than 2 million, Chinese, Russian - more than 1 million, Ukrainian - 680 thousand articles).

For initial access to the system there have been applied special terms - names of scientists and terms of targeted issues with the relevant articles on the resource that are created and edited by authors-experts. Along with scientists indices on the proposed methodology the network domains responsible to the authors are built, too. This aspect, in our view, adds importance to the proposed approach.

A sufficient amount of works and publications are dedicated to the research of subject areas as well as to the Wikipedia service that prove the relevance of the conducted studies [4]. The methods of building networks of co-authors, the definition of significant nodes of the network structure, research citations and appropriate buildings [5] are among them.

Also authors have studied the array of publications relating to the approaches to the assessment of citations and other aspects of the update, existence, filling, editing of the encyclopedic resource Wikipedia.

All the research on the design of subject domains and the assessments of links to certain publication or availability of links to articles in Wikipedia in scientific articles (in journals that are indexed by several scientometric system) is quite straightforward and relate only to a limited range of scientific fields, and definition "Wikipedia risks" of errors in scientific publications.

Based on the results of the processed data, we can assume the uniqueness of the proposed indices and value of the information that will be obtained by the computations to evaluate the level of certain data in the system of science popularization and accessibility of provided research information on specific issues.

The use of indices is appropriate in different directions of evaluation and analysis of scientific activity, can also act as an additional tool for decision making, forming educational programs etc.

Also using the received indices will contribute the encyclopedic resource development.

**The rule of Wikipedia Index computation**

The authors suggested the following rules for calculating Wiki-index of author's popularity. It is supposed that the references on the author are found in N Wikipedia articles.

Sorted by decreasing number of parameters that determine how many times author's name happens in bibliographic references of these articles we will denote as: $R_1$, $R_2$, ..., $R_N$.

Wiki-index of author's popularity ($WI$) corresponds to the maximum number of articles ($WH$) of Wikipedia, in which the number of references no more than the $WH$ value, which is multiplied by a certain integral function, which is not decreasing (eg, the square root is considered below) the N, that is:

$$WI = WH \times \sqrt{N} = \max(i : R_i > i) \times \sqrt{N}$$

Wiki-index of author popularity is ideologically close to the Hirsch index; however, it doesn't take into account the number of articles that refer to the author's article and citations to the work of the author and the number of articles from Wikipedia, which contain these data links. Another difference from the Hirsch index is the multiplication by a function of N, reflecting the consideration it provides greater popularity and the more spread of index values for different authors.

It should be noted that the level of popularity of the author must be attached to his subject domain on one hand in order to avoid false counting for homonyms, and on the other - to ensure completeness on subject area.

**Example:**

Let assume that the Wikipedia article with the highest number of references to author George Smith (in a given subject area) contains 100 references. The second - 20 documents, a third - 10, fourth – 5, fifth - 5, 4 more – only one link. So we have a number of values:

$R_1=100, R_2=20, R_3=10, R_4=5, R_5=5, R_6=1, R_7=1, R_8=1, R_9=1$

1 article contains the number of references least $R_1=100$;

2 articles contain the number of references least $R_2=20$;

3 articles contain the number of references least $R_3=10$;

4 articles contain the number of references least $R_5=4$.

5 articles contain the number of references least $R_5=5$.

There are no 6 articles that contain the number of references least 6.

In this case:

$N = 9, \ WH = 5,$

As follows, $WI = 5 \times \sqrt{9} = 15.$

**Algorithm**

In the process of the Wiki-index calculating there should be provided the procedure of Wikipedia resources scanning, corresponding to the subject area in which the author works. Accordingly, as "adverse product" of the Wiki-index computations, a model of the subject domain is being built, the model – is the network – nodes are concepts that represent articles from Wikipedia, and edges – are the hyperlinks between articles.

The process of the subject domain model of the author forming is possible in two ways:

- The use of Wikipedia dump database (not really relevant, but the link is available) by which the full range of all possible concepts and relationships. The advantage of this approach - completeness of information, disadvantage - possible loss of accuracy due consideration of homonyms, going beyond the subject area, considerable calculation time;

- The use of the principle of network services sounding (small sample volume of important contents of large information networks for technological reasons cannot be subjected to a complete scan). The advantage of this approach - getting accurate information strictly within a several subject domain, solvation of the homonyms problem and a short calculation time. The main drawback - the possible slight completeness, which may be assessed by additional experiments.

Authors chose the second approach for the Wiki-index computation while building its corresponding domain model chose the second approach, which was implemented as a software as a service.

**Formation of subject domain by sounding Wikipedia**

To implement calculation of Wiki-index authors considered the following algorithm to form subject domains according to Wikipedia, avoiding the effect of topic drift):

1. On the https://www.wikipedia.org/ in the search line the initial word is given, eg «**Albert Einstein**».
2. The search window opens. It contains information about conspet, according to the task on the Step1. The initial word/word combination is a graph vertex, which will be formed as the result of scanning.
3. All terms-concepts corresponding the hyperlinks on the chosen page, are added to the formed graph.

4. All the words/words combinations are the nodes of the graph. The edges to them are formed from the initial node.
5. The next transition is made by the first not involved hyperlink from the examining pages.
6. In text on the page to which the transition has been made the search of shortened researcher's name (eg, Einstein) or CAPTCHA (eg, **physics, relativity**) is to be carried out.
7. In case, if there is a shortened researcher's name or CAPTCHA is found, the transition to the Step 4 is made and accordingly from the node – word/word combination of the current search the new nodes are built.
8. If there is no word/word combination in the text – the given graph branch is considered to be built.
9. The next transition presumes pass to the page, which had been scanned – the word is not added as a graph node, and the feedback to the created node is formed.
10. All the operations under steps 4-9 repeat until the not involved hyperlinks, chosen from the page, are left. In another case the graph is considered to be built.

According to the suggested algorithm the data collection process in Wikipedia from the first node-notion is stopped when according to the algorithm transition to the new node is impossible (there are no more basic nodes for transition), so the "loop" is impossible.

**Calculation of the Wikipedia index of author's popularity**

To compute the Wiki-index it is necessary to make some changes to the suggested above algorithms, that is on the page, transition to which had been made by the hyperlink ($5^{th}$ Step of the algorithm), the search of author mentions in **Publications, References, Further Reading** sections is provided.

Herewith, the number of these mentions, which correlates values $R_i$, is counted. If $R_i = 0$, the article is not important, the concept is defined as the endnote and the transition to the Step 4 is provided. Of course, this rule narrows the scanning of Wikipedia pages list and results the completeness loss, though, as the real computations prove, has little effect on the overall results. Pages dedicated to the scientific concepts and those, which don't contain relevant publications, can be ignored – just skipped. Therewith, the time of Wikipedia target segment is significantly reduced.

As a result of the full network sounding, the sequence $R_1$, $R_2$, ..., $R_N$ is formed, which is used to calculate Wiki-index, according to the rules above.

**Experimental section**

**Calculation**

The represented algorithms were implemented as a software system, through which the subject domains models and Wiki-index are formed. Here are some examples of calculating Wiki-indices for three authors: Albert Einstein, Enrico Fermi, Benoit Mandelbrot.

Fig. 1 shows the program trace fragments, providing the Wikipedia sounding, and depict concepts to which the transition from initial concepts to concepts that include the author's name or the CAPTCHA.

| Albert_Einstein | Enrico_Fermi | Benoit_Mandelbrot |
| --- | --- | --- |
| **1: Albert_Einstein** | **1: Enrico_Fermi** | **1: Benoit_Mandelbrot** |
| SCI Links (1): 174 | SCI Links (1): 28 | SCI Links (1): 47 |
| 0 Rd +: Ulm | 0 Rd -: Rome | 0 Rd +: Mandelbrot_set |
| 1 Rd -: German_Empire | 1 Rd -: Chicago | 1 Rd -: Warsaw |
| 2 Rd -: Statelessness | 2 Rd -: Physics | 2 Rd -: Second_Polish_Republic |
| 3 Rd -: Switzerland | 3 Rd +: Leiden_University | 3 Rd -: Mathematics |
| 4 Rd -: Kingdom_of_Prussia | 4 Rd -: University_of_Florence | 4 Rd -: Aerodynamics |
| 5 Rd -: Free_State_of_Prussia | 5 Rd -: Columbia_University | 5 Rd -: Yale_University |
| 6 Rd -: Weimar_Republic | 6 Rd +: University_of_Chicago | 6 Rd -: IBM |
| 7 Rd -: Physics | 7 Rd -: Alma_mater | 7 Rd -: Alma_mater |
| 8 Rd -: Philosophy | 8 Rd -: Doctoral_advisor | 8 Rd -: University_of_Paris |
| 9 Rd -: Swiss_Patent_Office | 9 Rd +: Luigi_Puccianti | 9 Rd -: Eugene_Fama |
| 10 Rd -: Bern | 10 Rd +: Max_Born | 10 Rd +: Ken_Musgrave |
| 11 Rd -: University_of_Bern | 11 Rd -: Paul_Ehrenfest | 11 Rd -: Murad_Taqqu |
| 12 Rd -: University_of_Zurich | 12 Rd +: Harold_Agnew | 12 Rd +: Mandelbrot_set |
| 13 Rd +: ETH_Zurich | 13 Rd -: Edoardo_Amaldi | 13 Rd +: Chaos_theory |
| 14 Rd -: Kaiser_Wilhelm_Institute | 14 Rd +: Owen_Chamberlain | 14 Rd +: Fractal |
| 15 Rd -: German_Physical_Society | 15 Rd +: Geoffrey_Chew | 15 Rd -: Johannes_Kepler |
| 16 Rd +: Leiden_University | 16 Rd +: Jerome_Isaac_Friedman | 16 Rd +: Szolem_Mandelbrojt |

Fig.1. Wikipedia sounding program traces fragments

In Fig. 2 shows the Gephi visualization of domain model fragments that were obtained by sounding Wikipedia according to the above algorithm. The parameters of obtained networks (subject domain models); nodes-concepts of Wikipedia are following.

For a network that meets the model of authors' subject domain:

**Albert Einstein:**

- nodes – 718,

- Edges – 22111,

- The average degree of a node – 62,

- The graph diameter – 4,

- The average rate of clustering – 0, 26,

- the largest nodes:

| Consept | The node degree |
| --- | --- |
| Quantum_nonlocality | 188 |
| Alain_Aspect | 181 |
| Hermann_Weyl | 177 |
| Paul_Dirac | 174 |
| Electromagnetic_radiation | 174 |

| Isaac_Newton | 169 |
| Galileo_Galilei | 169 |
| Wolfgang_Pauli | 169 |
| General_relativity | 167 |
| Antimatter | 167 |

$WI = 12 \times 11,5 = 138$, ( 128 articles with the references, $WH = 12$ )

**Enrico Fermi:**

- Nodes – 605,

- Edges – 22079,

- The average degree of a node – 73,

- the graph diameter – 4,

- The average rate of clustering – 0,47,

- the largest nodes:

| Consept | The node degree |
|---|---|
| Enrico_Fermi | 440 |
| Nobelium | 206 |
| Transuranic_element | 206 |
| Particle_physics | 204 |
| Mendelevium | 204 |
| Einsteinium | 204 |
| Berkelium | 203 |
| Radioactive_decay | 195 |
| Radioactive | 190 |
| Particle_accelerators | 188 |

$WI = 7 \times 9,6 = 67$, (92 articles with the references, $WH = 7$)

**Benoit Mandelbrot:**

- nodes – 34,

- edges – 259,

- the average degree of a node – 15,2,

- the graph diameter – 3,

- the average rate of clustering – 0,52,

- the largest nodes:

| Conscept | The node degree |
|---|---|
| Benoit_Mandelbrot | 22 |
| Pattern | 20 |
| Chaos_theory | 18 |
| Patterns_in_nature | 18 |
| Hausdorff_dimension | 17 |
| Patterns | 16 |
| Fractal | 15 |
| Fractal_dimension | 15 |
| Fractal_geometry | 15 |
| Fractals | 15 |

$WI = 6 \times \sqrt{11} \approx 20$, (11 articles with the references, $WH = 6$)

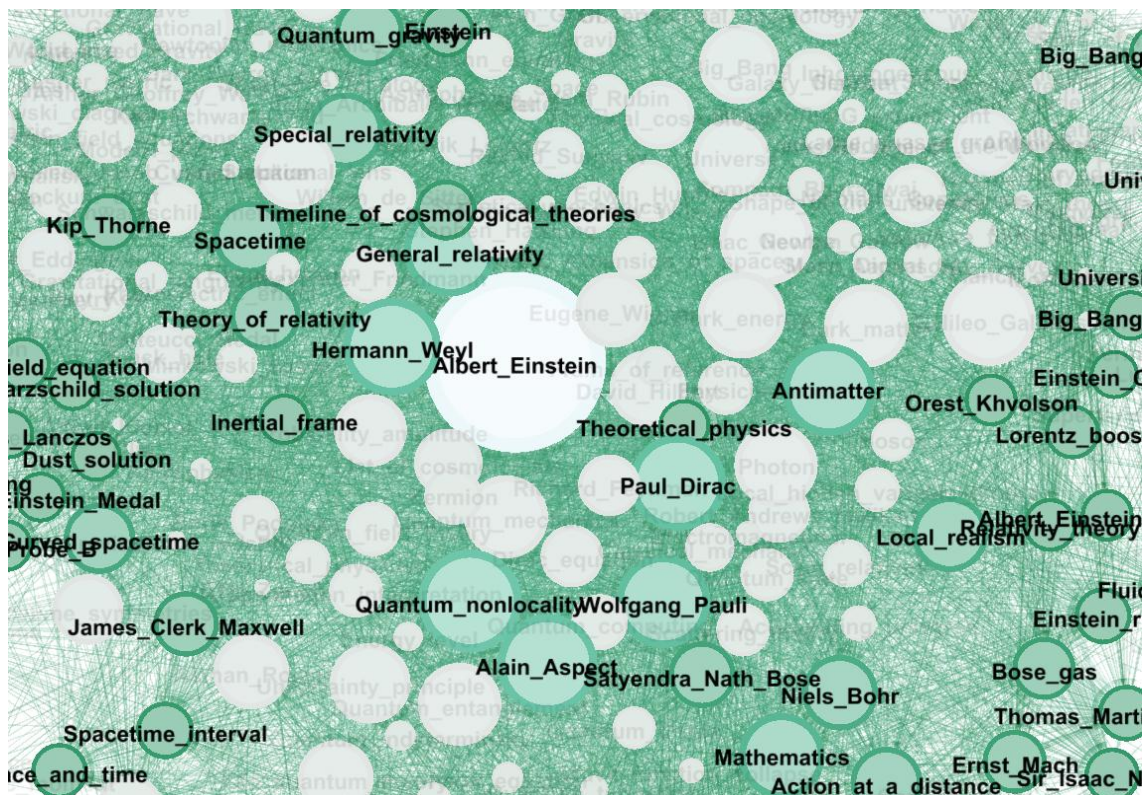

a)

b)

c)

Fig.2. Fragments of subject domains:

a) Albert Einstein, b) Enrico Fermi, c) Benoit Mandelbrot

There were provided comparisons of the results – Wiki-index, calculated on the research and the Hirsch-index, represented by the world's leading scientometric resources Scopus, Web Of Science and Google Scholar Citations.

Results are depicted in Table 1.

Table 1. Comparison of Wiki-indices values with the Hirsch-index (Scopus, Web Of Science та Google Scholar Citations)

| N | Scientist | Wiki-index | h-index Scopus | h-index Web Of Science | h-index Google Scholar Citations |
|---|---|---|---|---|---|
| 1. | Albert Einstein | 141 | 36 | 6 | 110 |
| 2. | Enrico Fermi | 67 | 26 | 1 | 49* |
| 3. | Benoit Mandelbrot | 20 | 31 | 36 | 90 |

*Profile missing, the value was calculated for: "e. fermi" according to the Google Scholar Calculator service

By comparison, we can see and estimate the role of information on research and publications on open-access resources in comparison with data that consider purely scientific information with a certain restrictions set.

**Conclusions**

As a result of calculations and proposed approaches tests to the formation of popular author index due to the presence of references to his/her work and references in the largest encyclopedic resource – Wikipedia, following conclusions can be made:

1. The principle of Wiki-index forming differs primarily from those, which currently is used in scientometrics with consideration of citation from not only scientific papers but popular service Wikipedia. This way the index of author's popularity within this service can be obtained. This is an importaint issue, considering the fact that Wikipedia is currently the largest and most popular encyclopedic resource.

2. There is suggested the technique of the Wiki-index quick calculation, which allows to realize computation as a separate service, and also automatically form the subject domain.

3. Due to the use and promotion of proposed indicies there can be a significant expansion of open access resources (available to be edited by Internet users).

4. Provided work may be continued by analyzing other resources and the formation of indicators to estimate and analyze the influence in a particular environment.

It can be underlined that the Wikipedia system, as Google Scholar Citations,

Considered before [6; 7], is practical from the point of view of access to information, doesn't allow signing in procedure and personal profile formation to get the access to the information, the access in unlimited.

It is also necessary to note a fundamental difference between the proposed approach of automatic subject domains models formation and those that already exist, based on direct participation of experts in selecting specific nodes and links. In cases, as it depicted in the work, the researcher uses only a small share of knowledge represented by the name of the scientist, his writing abbreviated names of several key terms, concepts to construct an appropriate network. After that, the program uses the knowledge that is implanted by Wikipedia articles' authors (editors), tags defined by internal hyperlinks. This way expert area is widely extended.

[1] Dobrov, B. V., Solovyov, V. D., Lukashevich, N. V., Ivanov, V. V. (2009), Ontologies and thesauri: models, tools, applications [Ontologii i tezaurusy: modeli, instrumenty, prilozheniia], Binom, Moscow, 173 p. (In Russian)

[2] Lande, D. V., Snarskii, A. A. (2014), The approach to the creation of terminological ontologies [Podkhod k sozdaniiu terminologicheskikh ontologii], Ontology engineering, No. 2 (12), pp. 83-91. (In Russian)

[3] Chanyshev O.G.(2008), Automatic construction of terminological knowledge base [Avtomaticheskoe postroenie terminologicheskoj bazy znanij], Proceedings of the 10th Russian's cientific conference «E-libraries: perspective methods, relevent collections» , RCDL'2008, Dubna, Russia, pp. 85-92.
(In Russian).

[4] Zareen Saba Syed, Tim Finin, Anupam Joshi. Wikipedia as an Ontology for Describing Documents, Proc. 2nd Int. Conf. on Weblogs and Social Media, AAAI Press, March 2008.,pp. 136-144.

[5] Lande, D.V., Snarskii, A. A., Bezsudnov, I. V. (2009), Internetika: Navigation in complex networks: models and algorithms [Internetika : navigatciia v slozhnykh setiakh : modeli i algoritmy], Librokom (Editorial URSS), Moscow, 264 p. (In Russian).

[6] Lande D.V., Andrushchenko V.B. (2016) Creation of the co-authorship network of law experts on the basis of exploration  of the Google Scholar Citations service [Pobudova mereg spivavtorstva fahivtsiv z yuryspudentsii za danymy  servisu Google Scholar Citations],  Legal informatics, 2016, № 1(46),  pp. 146-150. (In Ukrainian).

[7] Lande D.V. (2015) Creation of a domain model by probing Google Scholar citations. [Postroenie modeli predmetnoj oblasti putem zondirovaniya servisa Google Scholar citations] , Ontology of Designing, №3(17), pp.328-335 (In Russian).